# Assessing the association between pre-course metrics of student preparation and student performance in introductory statistics: Results from early data on simulation-based inference vs. non-simulation based inference

Nathan Tintle, Department of Mathematics and Statistics, Dordt College, Sioux Center, IA, USA

Jake Clark, Department of Biostatistics, University of Iowa, Iowa City, Iowa, USA

Karen Fischer, Department of Statistics, Texas A&M, College Station, TX, USA

Beth Chance, Department of Statistics, California Polytechnic University, San Luis Obispo, CA, USA

George Cobb, Department of Mathematics, Mount Holyoke College, South Hadley, MA, USA

Soma Roy, Department of Statistics, California Polytechnic University, San Luis Obispo, CA, USA

Todd Swanson, Department of Mathematics, Hope College, Holland, MI, USA

Jill VanderStoep, Department of Mathematics, Hope College, Holland, MI, USA

## Abstract

The recent simulation-based inference (SBI) movement in algebra-based introductory statistics courses (Stat 101) has provided preliminary evidence of improved student conceptual understanding and retention. However, little is known about whether these positive effects are preferentially distributed across types of students entering the course. We consider how two metrics of Stat 101 student preparation (pre-course performance on concept inventory and math ACT score) may or may not be associated with end of course student performance on conceptual inventories. Students across all preparation levels tended to show improvement in Stat 101, but more improvement was observed across all student preparation levels in early versions of a SBI course. Furthermore, students' gains tended to be similar regardless of whether students entered the course with more preparation or less. Recent data on a sample of students using a current version of an SBI course showed similar results, though direct comparison with non-SBI students was not possible. Overall, our analysis provides additional evidence that SBI curricula are effective at improving students' conceptual understanding of statistical ideas post-course regardless student preparation. Further work is needed to better understand nuances of student improvement based on other student demographics, prior coursework, as well as instructor and institutional variables.

## 1. Introduction

Students taking college-level, algebra-based introductory statistics courses (Stat 101) come to the course having a variety of different amounts of preparation. For example, some students have completed high school algebra or college algebra, some have completed high school statistics, and still others have completed high school pre-calculus, calculus or AP Statistics. A struggle when teaching Stat 101 is accommodating these varying levels of preparation appropriately, so that all students can learn effectively: strong students are not bored, nor are weaker students left behind, and all students improve on key measures of conceptual understanding.

Numerous studies have looked at variables related to student performance in statistics courses. Prior research has shown that prior mathematical skills, as measured by the ACT and a basic math skills test, were strong predictors of student performance (Johnson & Kuennen, 2006), while the type and order of college level mathematics courses taken was strongly associated with student performance in an introductory statistics course in a business school (Green, Stone, Zegeye, & Charles, 2009). Another study of student performance in business statistics suggested that both performance in an algebra course and college GPA were strong predictors of student performance in the course (Rochelle & Dotterweich, 2007). Two recent studies demonstrated that college GPA, as well as ACT, were both strong predictors of student performance in a general, multi-disciplinary introductory statistics course (Li, Uvah, & Amin, 2012; Wang, Tu, & Shieh, 2007). This finding is in line with other research which has indicated that poor mathematical training in high school leads to challenges learning statistics in college (Gnaldi, 2006). Another study, focusing on performance in an introductory statistics course for psychology students, found that age and an assessment of algebra skills were strong predictors of student performance (Lester, 2007). A recent multi-institutional study concluded that increased mathematical coursework in high school was associated with students later taking more and harder statistics courses in college, and performing better in those courses (Dupuis et al., 2011). Two studies which considered both mathematical competencies as well as student attitudes found that both were significant predictors of student performance (Cherney & Cooney, 2005; Silvia, Matteo, Francesca, & Caterina, 2008).

These studies examining student performance have generally focused on mathematical and quantitative reasoning (e.g., ACT score) and general measures of student performance (e.g., class grade and GPA), as these metrics are generally readily available. However, these studies are limited in at least three important ways. First, these studies mainly look at predicting students' end of course performance using prior predictors: they fail to consider what factors may be associated with students' *change* in statistical understanding throughout the course. For example, are students who end the course at a high level of conceptual understanding simply doing so because they started with a high level of understanding? Second, in these studies, student performance is generally measured using course grades, instead of using valid and reliable measures of students' conceptual understanding of statistics in the course. Finally, these studies are primarily focused on looking at student performance in a single course, and not drawing comparisons across curricula. In particular, given the recent gain in popularity of simulation-based inference (SBI) as an approach for teaching introductory statistics (N. L. Tintle et al., 2015; N. Tintle, VanderStoep, Holmes, Quisenberry, & Swanson, 2011), an important and unaddressed question is how student abilities and background may be associated with student performance differently in SBI curricula vs. traditional curricula.

In this paper, we address these limitations by focusing primarily on how students' conceptual understanding of statistics changes from the beginning of the course to the end using a valid and reliable instrument. First, we explore how students' growth in conceptual understanding may be different depending on students' mathematical or statistical abilities when they enter the course as measured by ACT score, a statistics concepts pre-test or their college GPA. For two institutions, we have data both before and after a switch to an early-SBI curriculum, and compare student performance both across ability/preparation groups as well as across curricula. We examine overall performance as well as performance within different subscales. Second, for a larger set of institutions using an SBI curriculum, we evaluate whether students at all ability/preparation levels show similar improvement in conceptual understanding overall, and within subscales, in light of recent results showing improved conceptual performance among students in simulation-based courses (Chance & Mcgaughey, 2014; N. L. Tintle et al., 2014; N. Tintle, Topliff, VanderStoep, Holmes, & Swanson, 2012; N. Tintle et al., 2011).

## 2. Methods

In our analysis we considered students using three different curricula: (1) Students using a traditional (consensus curriculum) Stat 101 textbook (denoted "consensus"), (2) Students using a preliminary version of a simulation-based inference curriculum (denoted "early-SBI") and (3) Students using a full version of a simulation-based inference curriculum (denoted "SBI"). The following three sections briefly summarize each curriculum, while Section 2.4 explains which students and assessments were used with each curriculum. We now briefly describe each curriculum.

### 2.1. Consensus curriculum (consensus)

For more than a decade Stat 101 has had a generally accepted consensus curriculum focused on the normal distribution and its derivatives, such as the *t*-distribution, for conducting statistical inference (Malone, Gabrosek, Curtiss, & Race, 2010; Scheaffer, 1997). Students at two institutions (both small, Midwestern liberal arts colleges) used textbooks following a consensus curriculum in 2007 and spring 2011 before switching to SBI see (N. L. Tintle et al., 2014; N. Tintle et al., 2012, 2011) additional details on this curriculum and its implementation.

### 2.2. Early simulation-based inference curriculum (early-SBI)

The same two institutions which used the consensus curriculum switched to an early version of an SBI curriculum in subsequent years (2009 and the 2011-12 academic year, respectively). In this curriculum, unlike the consensus curriculum, formal statistical inference is introduced to students early in the course, motivated through tactile and computer-aided simulations in contrast to formal, mathematical representations of sampling distributions. Notably, student profiles were similar before and after the switch to the early-SBI curriculum, and a subset of instructors were similar before and after the switch. See other manuscripts for a full description of the early version of this curriculum, its implementation and other student and instructor details (N. L. Tintle et al., 2013; N. Tintle et al., 2012, 2011).

### 2.3. Simulation-based inference curriculum (SBI)

Significant revisions were made to the early SBI curriculum in recent years, primarily with regards to the ordering of topics, while maintaining the focus on SBI. For example, in the early-SBI curriculum simulation-based methods were covered in the first half of the course and theory-based methods in the latter half, while in the new approach chapters were based around data contexts, meaning the simulation-based and theory-based methods were presented back-to-back for each type of data analysis. This curriculum was utilized at numerous institutions in the 2013-2014 academic year, and was recently published in its first edition, which is only very modestly different from the version used in 2013-14 (N. Tintle et al., 2016).

## 2.4. Samples and assessments

The samples and assessments used in this analysis have been described in detail elsewhere (Consensus and early-SBI: (N. L. Tintle et al., 2014; N. Tintle et al., 2012, 2011); SBI (Chance, Wong, & Tintle, n.d.)), we provide a brief overview here. The consensus curriculum sample consists of 289 students from two institutions, covering two separate semesters and multiple instructor-sections of approximately 30 students per instructor-section. All students completed the Comprehensive Assessment of Outcomes in Statistics (CAOS; (del Mas, Garfield, Ooms, & Chance, 2007)) during the first week of classes and again during the final week of class. Response rates were over 85% per section. The early-SBI sample consists of 366 students from the same two institutions as the consensus sample, with similar demographic characteristics and obtained in semesters shortly after the consensus sample data was obtained. The assessment was taken online and outside of class, with a small incentive (homework points) for completion, but not for performance.

The SBI sample consists of 1078 students from 34 instructor sections across 13 institutions including one community college, one private university, two high school AP statistics courses, four private liberal arts colleges and five large, public universities. These instructors administered a modified version of the CAOS test which exchanged some questions that most students tend to get right or get wrong with alternative formulations or alternative questions altogether. See Chance et al. (2017) for details. Test administration varied a bit between sections, but generally followed the same pattern as described for the consensus and early SBI samples. All data collection was approved by the ABC College Institutional Review Board. Composite ACT scores were gathered from the institutional research office for a single institution, while college GPA was gathered via self-report from all students in the SBI sample.

## 2.5. Statistical analysis

Statistical analyses examine pre- to post-course changes in test performance either on the test as a whole, or on subscales of the test covering similar topics (see N. Tintle et al., 2011 for details). In general, analyses examine either statistical significance between pre-test and post-test (paired *t*-tests) or whether change scores (post-test minus pre-test) are significantly different across subgroups (e.g., SBI vs. non-SBI curricula). In linear models testing for differences in change scores across consensus and early-SBI curricula, institution is adjusted for as a covariate (Change (Post-test minus pre-test) = Curriculum +Institution). A significance level of 0.05 is used throughout.

# 3. Results

## 3.1. Consensus curriculum vs. early-SBI: Overall results

On average, students in the early-SBI curriculum showed similar performance on the CAOS pre-test to students taking the consensus curriculum (Consensus mean: 47.7% vs. SBI mean: 44.6%) when data was combined across both institutions. The distribution of pre-test scores approximately followed those of nationally representative samples (delMas et al. 2007). We separated students into three groups of approximately equal size (tertile split) based on pre-test scores (40% or lower correct (Low performance), between 40 and 50% correct (Middle performance) or 50% or more correct (High performance)). Table 1 provides the full details of the distribution of pre-test and post-test scores between the two curricula.

**Table 1.** Pre- and post-course CAOS scores stratified by pre-course performance and curriculum

| Pre-test score group | Curriculum | Pre-test mean % correct (SD[1]) | Post-test mean % correct (SD[1]) | Change in mean % correct (SD[1])[2] | Difference in mean change by curriculum (SE)[3] |
|---|---|---|---|---|---|
| Low (≤40%) | Consensus (n=80) | 35.2 (4.9) | 48.1 (8.8) | 12.9 (9.4)*** | 1.9 (1.4) |
|  | Early-SBI (n=141) | 35.2 (4.8) | 50.1 (9.8) | 14.9 (10.6)*** |  |
| Middle (Between 40 and 50%) | Consensus (n=77) | 45.1 (2.1) | 52.0 (10.2) | 6.9 (10.2)*** | 3.1 (1.5)* |
|  | Early-SBI (n=108) | 44.9 (2.0) | 54.9 (10.8) | 10.0 (10.4)*** |  |
| High (≥50%) | Consensus (n=129) | 57.1 (6.8) | 62.3 (6.8) | 5.2 (9.1)*** | 2.1 (1.3) |
|  | Early-SBI (n=117) | 55.8 (5.6) | 63.0 (11.3) | 7.2 (10.5)*** |  |
| Overall | Consensus (n=289) | 47.7 (10.7) | 55.5 (11.8) | 7.8 (10.0)*** | 3.2 (0.01)*** |
|  | Early-SBI (n=366) | 44.6 (9.7) | 55.6 (11.9) | 11.0 (11.0)*** |  |

*p<0.05; **p<0.01; ***p<0.001

1. SDs are SDs of student test scores (pre-test or post-test) or SD of change in student test scores.

2. Significance is indicated by asterisks and reported based on results from paired *t*-tests comparing the pre-test and post-test scores

3. From a linear model predicting the change in score by curriculum and adjusted for institution (Change = Curriculum + Institution). Institution was not significant in any of the four models (p-values of 0.62-low; 0.22- middle; 0.72- high; 0.38- overall). SE is the SE of the Curriculum term in the linear model.

Table 1 shows that all three pre-course performance groups performed better using the early-SBI curriculum (between 2 to 3 percentage points), but only the middle performing group's performance was statistically significant after adjusting for institutional differences. Table 2 shows results when students were stratified by their ACT scores (again, using an approximate tertile split for this sample). As with Table 1, the Early-SBI curriculum shows improved performance in all three groups compared to the consensus curriculum in this analysis, however in this case the results were significant in each of the three groups. Figure 1 summarizes this data visually illustrating gains for all students, but larger gains for the early-SBI students across ACT score groups.

**Table 2.** Pre- and post-course CAOS scores stratified by ACT score and curriculum[1]

| ACT Group | Curriculum | Pre-test mean % correct (SD[2]) | Post-test mean % correct (SD[2]) | Change in mean % correct (SD[2])[3] | Difference in mean change by curriculum[4] |
|---|---|---|---|---|---|
| Low ($\leq$22) | Consensus (n=21) | 41.7 (10.3) | 46.3 (10.1) | 4.0 (11.7) | 8.2 (2.9)*** |
| | Early-SBI (n=55) | 42.7 (10.1) | 54.9 (11.9) | 12.2 (10.5)*** | |
| Middle (23-26) | Consensus (n=34) | 46.0 (8.2) | 52.4 (10.3) | 6.5 (9.2)*** | 4.7 (2.1)* |
| | Early-SBI (n=48) | 44.0 (10.0) | 55.1 (10.8) | 11.2 (11.4)*** | |
| High ($\geq$27) | Consensus (n=36) | 51.3 (7.7) | 57.1 (7.7) | 5.8 (9.2)** | 6.0 (2.4)* |
| | Early-SBI (n=49) | 47.8 (9.8) | 59.5 (12.0) | 11.8 (10.1)*** | |
| Overall | Consensus (n=91) | 46.4 (9.3) | 52.0 (11.0) | 5.6 (9.8) | 6.0 (1.4)*** |
| | Early-SBI (n=152) | 44.9 (10.1) | 56.5 (11.6) | 11.6 (10.7) | |

1. Only for students with ACT scores available (all students with available ACT scores were from one of the two colleges evaluated in Table 1)
2. SDs are SDs of test scores or change in test scores
3. From a paired *t*-test comparing the pre-test and post-test scores
4. These values indicate how different the two curricula are with regards to changing student scores and are estimated from a linear model predicting Change in score by Curriculum). For example, 8.2 means that the Early-SBI curriculum shows an improvement in percent correct which is 8.2 percentage points higher than the Consensus curriculum. A test to see whether there was evidence that the difference in mean changes were different by ACT group (e.g., 8.2 vs. 4.7 vs. 6.0) did not yield evidence of a significant (p=0.15; ANOVA comparison of whether a model predicting post-test scores by pre-test, curriculum used and ACT score group was significantly different than a model which predicted post-test scores by pre-test and curriculum only).

**Figure 1.** Student performance on CAOS stratified by ACT score and curriculum

**Figure 1a**

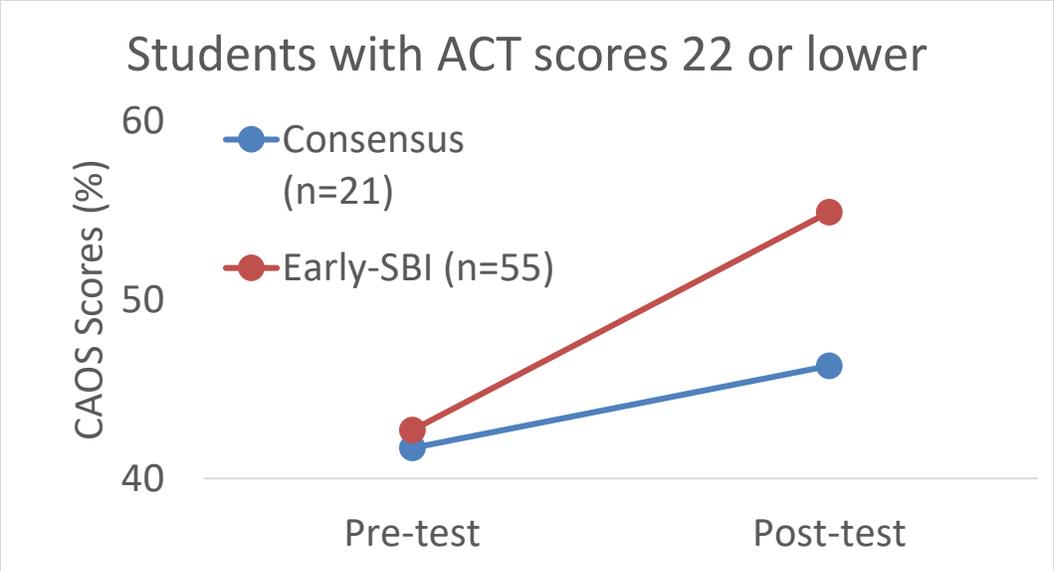

**Figure 1b**

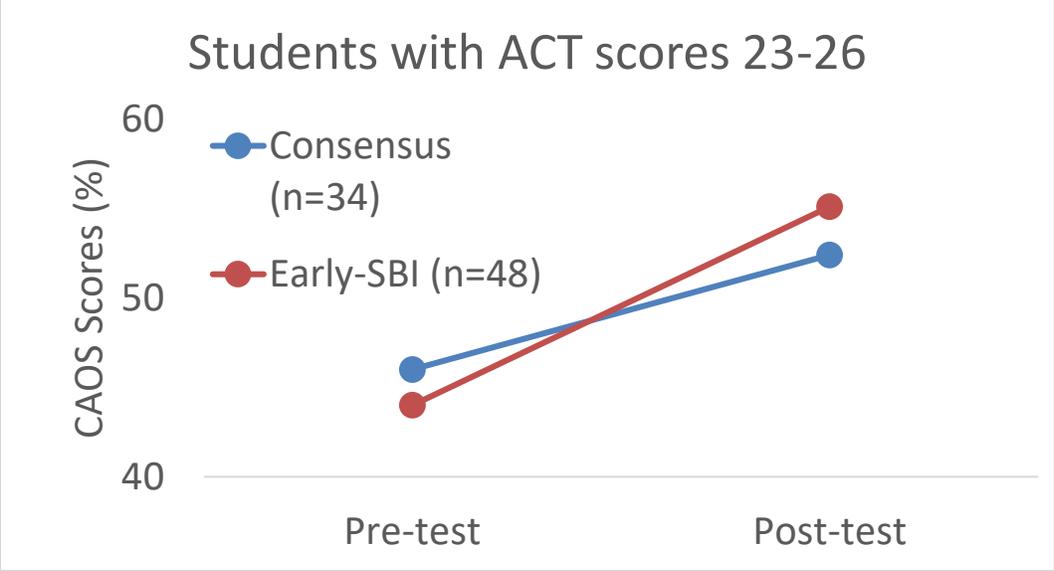

**Figure 1c**

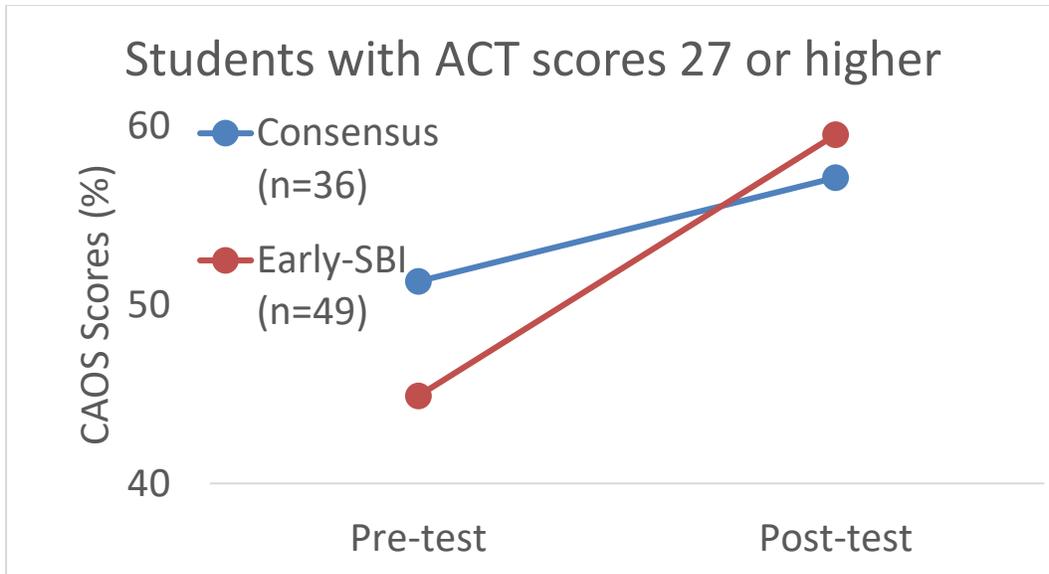

### 3.2. Consensus curriculum vs. early-SBI: Results by subscale

Table 3 shows the change in percentage correct from pre-test to post-test for each of the nine individual subscales of the CAOS test (delMas et al. 2007), by curriculum, for students in the lowest pre-course performance group (getting 40% or fewer questions correct on the pre-test). Across the nine subscales of the CAOS test, three subscales showed significant improvement: data collection and design, tests of significance, and probability. After adjusting for institution, these lower performing students showed an approximately 8.4 percentage point improvement on tests of significance, 9.4 percentage point improvement on data collection and design issues, and 15.8 percentage point improvement on probability topics. Data collection and design showed similar positive impact of the early-SBI curriculum for students with medium (40-50%) and high (50%+) performance on the pre-test (11.5 point curricular impact and 7.7 point curricular impact), though the improvement was only statistically significant for the medium group. Probability and simulation also only showed statistically significant curricular impact for the medium performing group (14.4 impact), with the highest group showing a 2.5 point improvement from the early-SBI curriculum. Tests of significance also more improvement for early SBI students relative to the consensus curriculum (6.5 points and 8.2 points better, respectively, for medium and high students), though this was not statistically significant for the medium students. The descriptive statistics subscale students showed significantly less improvement for the medium students using the early-SBI curriculum as compared to the consensus curriculum (10 point decline). Appendix Tables A and B give full results for students in the Middle and High performing pre-test performance groups.

**Table 3.** Pre- and post-course CAOS scores by subscale and curriculum – Low performing students

| Subscale (# of items) | Curriculum | Pre-test mean % correct (SD[1]) | Post-test mean % correct (SD[1]) | Change in mean % correct (SD[1]) | Difference in mean change by curriculum (SE)[2] |
|---|---|---|---|---|---|
| | | | | | |
| Graphical Representations (9 items) | Consensus (n=80) | 44.2 (15.6) | 58.5 (17.7) | 14.3 (20.9) | 0.2 (2.9) |
| | Early-SBI (n=141) | 42.1 (16.0) | 56.4 (17.1) | 14.3 (20.2) | |
| Box plots (4 items) | Consensus (n=80) | 21.9 (19.2) | 37.5 (21.0) | 15.6 (27.1) | -4.6 (3.9) |
| | Early-SBI (n=141) | 27.3 (20.0) | 37.4 (22.3) | 10.1 (28.5) | |
| Data collection and design (4 items) | Consensus (n=80) | 23.4 (21.3) | 27.9 (21.5) | 4.2 (25.1) | 9.4 (4.0)* |
| | Early-SBI (n=141) | 21.7 (19.5) | 36.9 (26.6) | 15.1 (31.0) | |
| Descriptive statistics (3 items) | Consensus (n=80) | 32.8 (19.7) | 50.5 (24.8) | 17.8 (27.0) | -6.7 (3.5) |
| | Early-SBI (n=141) | 32.9 (18.2) | 44.0 (22.3) | 11.1 (23.8) | |
| Tests of significance (6 items) | Consensus (n=80) | 36.9 (19.1) | 54.0 (19.5) | 17.1 (28.3) | 8.4 (3.8)* |
| | Early-SBI (n=141) | 40.4 (18.2) | 66.3 (17.9) | 25.9 (26.4) | |
| Bivariate relationships (3 items) | Consensus (n=80) | 49.4 (22.8) | 55.3 (19.4) | 5.9 (27.3) | 5.3 (4.2) |
| | Early-SBI (n=141) | 45.6 (23.8) | 56.7 (21.7) | 11.2 (30.7) | |
| Confidence Intervals (4 items) | Consensus (n=80) | 35.0 (20.9) | 48.8 (23.2) | 13.8 (31.0) | 1.8 (4.2) |
| | Early-SBI (n=141) | 34.4 (21.5) | 49.6 (25.5) | 15.2 (28.6) | |
| Sampling Variability (5 items) | Consensus (n=80) | 28.8 (26.4) | 36.7 (28.9) | 7.9 (39.8) | -4.3 (5.3) |
| | Early-SBI (n=141) | 30.0 (28.0) | 33.6 (26.6) | 3.5 (36.2) | |
| Probability/Simulation (2 items) | Consensus (n=80) | 21.3 (26.1) | 31.3 (28.0) | 10.0 (36.8) | 15.8 (5.6)** |
| | Early-SBI (n=141) | 18.8 (25.7) | 46.0 (32.2) | 27.3 (41.2) | |

1. SDs are of test scores or change in test scores
2. From a linear model predicting the difference in scores by curriculum and institution. SE is estimated as part of the linear model. Values are interpreted as the adjusted difference in curriculum with regards to change in percent correct. For example, 15.8 means that the Early-SBI students improved 15.8 percentage points more than Consensus curriculum students.

### 3.3. SBI curriculum results: Overall and by subscale

Table 4 illustrates the overall performance questions of SBI students on the modified CAOS across multiple institutions when stratified by pre-test score or by self-reported GPA. As was done earlier, pre-test concept score groups were created by breaking the sample into three approximately equal-size groups (tertiles) based on percentage correct (Low: 40% or less correct (n=291); Middle: Between 40 and 55% correct (n=422); High: 55% or better (n=365)). College GPA groups were created by creating a group with GPAs of B or worse (Low: Less than or equal to 3.0; n=193), GPAs between B+ and A- (Middle: Between 3.0 and 3.7; n=654), and GPAs in the A range (High: 3.7 or above; n=231).

Table 4. Pre- and post-course concept scores stratified by pre-course performance and GPA among SBI students in 2013-2014 (n=1078)

| How grouped | Grouping | Pre-test Mean (SD[1]) | Post-test Mean (SD[1]) | Change Mean (95% CI)[1] |
|---|---|---|---|---|
| Pre-test concept score | Low | 35.0 (5.0) | 50.2 (12.0) | 15.2 (113.7, 16.7)*** |
| | Middle | 48.1 (3.8) | 56.2 (12.1) | 8.1 (7.0, 9.2)*** |
| | High | 64.1 (7.3) | 68.1 (12.8) | 4.0 (2.9, 5.1)*** |
| Self-reported college GPA | Low | 45.6 (12.3) | 52.9 (13.1) | 7.3 (5.6, 9.0)*** |
| | Middle | 50.0 (12.0) | 58.1 (13.6) | 8.1 (7.2, 9.1)*** |
| | High | 53.8 (13.6) | 64.9 (14.7) | 11.1 (9.5, 12.7)*** |
| | Overall | 50.0 (12.6) | 58.6 (14.3) | 8.6 (7.9, 9.4)*** |

1. From a paired *t*-test comparing the pre-test and post-test scores

All subgroups show significant improvement. Improvement is largest for the least well prepared group when stratifying by pre-test score (Low (15.2) vs. Middle (8.1): p<0.001; Low (15.2) vs. High (4.0): p<0.001), and largest for students with the largest GPAs when stratifying by GPA (Low (7.3) vs. Middle (8.1): p=0.41; Low (7.3) vs. High (11.1): p=0.002).

Table 5 illustrates improvement scores within groups of similar conceptual questions. For the group of students with the lowest pre-test scores, significant improvement is seen within all seven conceptual areas. When stratifying by GPA, four of the seven groups of questions showed significant improvement for students with the lowest GPAs.

Table 5. Pre- and post-course conceptual understanding by subscale – Lower performing students (either based on pre-test group or GPA) students in 2013-2014

| Subscale | Grouping | Pre-test Mean (SD[1]) | Post-test Mean (SD[1]) | Change Mean (95% CI)[1] |
|---|---|---|---|---|
| Graphical Representations | Pre-test | 30.5 (18.4) | 46.5 (20.9) | 15.8 (12.9, 18.7)*** |
| | GPA | 44.0 (25.1) | 51.3 (25.0) | 6.5 (2.7, 10.3)*** |
| Data collection and design | Pre-test | 50.9 (21.2) | 56.6 (23.9) | 5.2 (1.5, 8.9)** |
| | GPA | 62.6 (23.5) | 60.4 (25.9) | -2.6 (-7.5, 2.3) |
| Descriptive statistics | Pre-test | 17.2 (27.2) | 31.8 (35.1) | 14.9 (9.8, 20.0)*** |
| | GPA | 31.3 (34.8) | 36.1 (33.4) | 4.4 (-1.9, 10.8) |
| Tests of significance | Pre-test | 40.7 (13.6) | 57.5 (17.4) | 16.6 (14.0, 19.2)*** |
| | GPA | 49.1 (17.0) | 60.0 (17.3) | 10.7 (7.6, 13.9)*** |
| Confidence Intervals | Pre-test | 33.4 (16.2) | 50.2 (22.7) | 17.1 (14.1, 20.2)*** |
| | GPA | 40.0 (18.5) | 50.4 (23.9) | 10.7 (7.0, 14.5)*** |
| Sampling Variability | Pre-test | 28.0 (29.3) | 40.6 (34.8) | 12.6 (7.3, 17.9)*** |
| | GPA | 44.8 (35.3) | 44.4 (39.2) | -0.4 (-7.4, 6.3) |
| Probability/Simulation | Pre-test | 19.9 (28.4) | 38.7 (34.9) | 18.2 (12.9, 23.4)*** |
| | GPA | 29.8 (31.2) | 41.9 (36.4) | 10.9 (4.8, 17.0)*** |

1. From a paired *t*-test comparing the pre-test and post-test scores

Similar patterns are observed for students in typical and more well-prepared groups as well (see Appendix Tables C and D). In particular, the middle performing group showed improvement on 5 of 7 scales when stratifying by either pre-test or GPA, while the highest group showed improvement on 3 of 7 scales when stratifying by pretest and 7 of 7 when stratifying by GPA.

## 4. Discussion

Results from early implementations of SBI curricula for Stat 101 have shown promising results, leading to better post-course performance on conceptual tests and better student retention post course. When stratifying students based on pre-course conceptual test performance, ACT score or self-reported GPA, all groups of students showed significant improvement in post-course understanding with some groups showing significantly greater improvement with SBI than with the consensus curriculum. Improvement among students using an early-SBI curriculum was greatest in areas related to data collection and design, tests of significance, and probability/simulation---all areas emphasized by the curriculum. A later version of the curriculum showed significant improvement among all topics, with largest improvements in tests of significance, confidence intervals, and probability/simulation.

In particular, we have demonstrated that the SBI curriculum considered here appears to work well for students across a range of preparation levels- regardless of prior statistical abilities, mathematical abilities, or general academic performance (college GPA). This provides a strong case that the accessible, tactile, and conceptual approach at the core of SBI is helping to level the playing field in introductory statistics courses, not simply making the top students better, nor is it only benefiting weaker students. Minor decreases in performance in descriptive statistics have been noted in early versions of the SBI curriculum (Tintle et al. 2011), but have gone away in assessment of more recent SBI curriculum (Chance et al. 2017; Tintle et al. 2014).

Some limitations of our analysis are worth noting. First, we note that the later version of the SBI curriculum was not compared against non-SBI curriculum, limiting the conclusions that can be drawn from the analysis. Results generally show similar trends as in early version of the curriculum, but further work is necessary to make comparisons across student preparation levels and curricula. Second, we note that tests were taken outside of class and for completion credit. Future work is needed to assess the impact of environment or different incentives on student performance on conceptual assessments in introductory statistics. Third, limited information was available on courses students might have taken prior to the course (e.g., calculus, AP statistics, etc.). Future work is needed to capture this information and assess it's ability to potentially better explain student performance in Stat 101 courses.

It is worth noting that although, in places, we focused on pre-test performance to stratify the sample, we recognize the impact of regression to the mean when comparing pre-test to post-test scores. With this in mind we looked at alternative stratification approaches including: ACT score and self-reported college GPA. Broad results are generally similar, showing that students across preparation levels show improvement in statistical thinking and reasoning. However, further work could consider different prior tests of statistical reasoning to classify students, as well as considering other demographic variables (e.g., race; socio-economic status) and their association with student performance in simulation-based inference curricula. Further, we note that using tertiles to split students based on ACT score yielded only 33% of the sample in the lowest ACT group (ACT<22), whereas in national samples 63% of students have ACT scores below 22 (ACT, 2017). Additional exploration of samples, especially among very academically under-performing students is warranted. Finally, another important area for further work is the consideration of student attitudes, especially among weaker students and how this impacts students' growth in conceptual understanding of statistical concepts.

## 5. References (after removing articles by the authors)

**Author contact information**

# Appendix

**Table A.** Pre- and post-course conceptual understanding by subscale – Middle performing students

| Group | Curriculum | Pre-test mean % correct (SD) | Post-test mean % correct (SD) | Change in mean % correct (SD) | Difference in mean change by curriculum (SE)[1] |
|---|---|---|---|---|---|
| Graphical Representations | Consensus (n=77) | 59.5 (15.3) | 64.6 (19.0) | 5.2 (1.90) | 3.0 (2.9) |
| | Early-SBI (n=108) | 57.0 (16.7) | 65.2 (15.3) | 8.2 (19.8) | |
| Box plots | Consensus (n=77) | 29.9 (19.0) | 36.4 (24.5) | 6.5 (32.3) | 1.3 (5.0) |
| | Early-SBI (n=108) | 34.3 (21.8) | 41.7 (27.6) | 7.4 (34.7) | |
| Data collection and design | Consensus (n=77) | 30.7 (19.3) | 32.0 (21.9) | 1.3 (26.7) | 11.5 (4.5)* |
| | Early-SBI (n=108) | 27.5 (19.8) | 40.7 (29.3) | 13.3 (32.5) | |
| Descriptive statistics | Consensus (n=77) | 43.9 (20.8) | 58.2 (23.9) | 14.3 (25.3) | -10.0 (3.8)* |
| | Early-SBI (n=108) | 45.0 (18.8) | 49.6 (23.4) | 4.6 (25.7) | |
| Tests of significance | Consensus (n=77) | 47.4 (20.2) | 57.1 (21.7) | 9.7 (26.8) | 6.5 (4.1) |
| | Early-SBI (n=108) | 48.9 (18.7) | 65.6 (19.4) | 16.7 (27.2) | |
| Bivariate relationships | Consensus (n=77) | 55.5 (18.0) | 60.4 (20.4) | 4.9 (25.0) | -1.4 (3.8) |
| | Early-SBI (n=108) | 59.5 (19.2) | 62.3 (19.5) | 2.8 (25.2) | |
| Confidence Intervals | Consensus (n=77) | 41.6 (18.0) | 46.1 (21.5) | 4.5 (28.9) | 7.1 (4.6) |
| | Early-SBI (n=108) | 39.8 (20.2) | 51.2 (25.0) | 11.3 (31.6) | |
| Sampling Variability | Consensus (n=77) | 36.4 (31.1) | 39.4 (27.4) | 3.0 (39.8) | 6.7 (5.7) |
| | Early-SBI (n=108) | 28.1 (25.4) | 36.7 (29.9) | 8.6 (36.8) | |
| Probability/Simulation | Consensus (n=77) | 27.9 (28.7) | 39.6 (30.7) | 11.7 (40.5) | 14.4 (6.8)* |
| | Early-SBI (n=108) | 31.5 (31.8) | 56.9 (32.4) | 25.5 (47.6) | |

[1] From a linear model predicting the difference in scores by curriculum and institution.

**Table B.** Pre- and post-course conceptual understanding by subscale – High performing students

| Group | Curriculum | Pre-test mean % correct (SD) | Post-test mean % correct (SD) | Change in mean % correct (SD) | Difference in mean change by curriculum (SE)[1] |
|---|---|---|---|---|---|
| Graphical Representations | Consensus (n=129) | 73.6 (14.1) | 74.2 (16.7) | 0.7 (17.5) | 2.3 (2.3) |
| | Early-SBI (n=117) | 70.4 (15.3) | 72.6 (14.6) | 2.3 (18.7) | |
| Box plots | Consensus (n=129) | 45.2 (24.0) | 49.4 (25.5) | 4.3 (24.0) | 2.4 (3.5) |
| | Early-SBI (n=117) | 42.1 (23.4) | 4.3 (24.0) | 7.3 (30.0) | |
| Data collection and design | Consensus (n=129) | 36.2 (23.9) | 42.1 (27.5) | 5.9 (29.9) | 7.7 (4.1) |
| | Early-SBI (n=117) | 35.3 (22.4) | 49.6 (32.3) | 14.2 (32.8) | |
| Descriptive statistics | Consensus (n=129) | 62.6 (21.0) | 71.2 (19.8) | 8.5 (22.6) | -3.8 (2.9) |
| | Early-SBI (n=117) | 61.2 (20.7) | 65.1 (23.2) | 3.9 (22.6) | |
| Tests of significance | Consensus (n=129) | 57.6 (20.6) | 68.0 (20.7) | 10.3 (25.4) | 8.2 (3.4)* |
| | Early-SBI (n=117) | 57.0 (19.2) | 75.9 (19.6) | 18.9 (26.8) | |
| Bivariate relationships | Consensus (n=129) | 64.1 (17.4) | 66.1 (17.3) | 1.9 (22.7) | -2.1 (3.0) |
| | Early-SBI (n=117) | 63.9 (17.2) | 64.3 (19.2) | 0.4 (23.2) | |
| Confidence Intervals | Consensus (n=129) | 45.9 (24.6) | 54.5 (23.9) | 8.5 (29.7) | 6.2 (4.1) |
| | Early-SBI (n=117) | 45.3 (23.0) | 59.8 (26.0) | 14.5 (33.4) | |
| Sampling Variability | Consensus (n=129) | 47.0 (24.5) | 50.4 (26.1) | 3.3 (35.1) | -7.1 (4.5) |
| | Early-SBI (n=117) | 46.7 (26.3) | 43.0 (26.3) | -3.7 (35.0) | |
| Probability/Simulation | Consensus (n=129) | 45.7 (35.4) | 51.2 (33.9) | 5.4 (38.1) | 2.5 (5.1) |
| | Early-SBI (n=117) | 49.1 (32.2) | 56.4 (32.5) | 7.3 (40.6) | |

[1] From a linear model predicting the difference in scores by curriculum and institution.

**Table C.** Pre- and post-course conceptual understanding by subscale – Middle performing students (either based on pre-test group or GPA) students in 2013-2014

| Subscale | Grouping | Pre-test Mean (SD) | Post-test Mean (SD) | Change Mean (95%CI)[1] |
|---|---|---|---|---|
| Graphical Representations | Pre-test | 46.9 (21.6) | 54.5 (22.1) | 7.1 (4.4, 9.8)*** |
| | GPA | 51.0 (25.4) | 57.7 (22.9) | 6.4 (4.4, 8.4)*** |
| Data collection and design | Pre-test | 62.9 (21.3) | 62.9 (24.6) | -0.3 (-3.4, 2.7) |
| | GPA | 62.9 (23.0) | 63.2 (24.6) | -0.0 (-2.4, 2.4) |
| Descriptive statistics | Pre-test | 30.9 (33.2) | 41.2 (37.0) | 10.8 (6.3, 15.4)*** |
| | GPA | 34.2 (36.4) | 43.8 (38.1) | 9.7 (6.1, 13.3)*** |
| Tests of significance | Pre-test | 53.2 (13.6) | 63.6 (16.9) | 10.5 (8.5, 12.6)*** |
| | GPA | 53.8 (15.9) | 64.6 (17.8) | 10.8 (9.1, 12.4)*** |
| Confidence Intervals | Pre-test | 43.3 (19.3) | 54.0 (22.9) | 11.0 (8.3, 13.6)*** |
| | GPA | 44.7 (19.7) | 54.9 (22.0) | 10.3 (8.2, 12.4)*** |
| Sampling Variability | Pre-test | 46.4 (33.0) | 48.0 (36.1) | 1.3 (-3.2, 5.7) |
| | GPA | 49.2 (34.8) | 53.1 (35.9) | 3.2 (-0.1, 6.5) |
| Probability/Simulation | Pre-test | 29.7 (30.8) | 40.0 (35.9) | 10.3 (5.8, 14.8)*** |
| | GPA | 32.7 (33.4) | 43.5 (36.2) | 10.6 (7.1, 14.2)*** |

1. From a paired *t*-test comparing the pre-test and post-test scores

**Table D.** Pre- and post-course conceptual understanding by subscale – High performing students (either based on pre-test group or GPA) students in 2013-2014

| Subscale | Grouping | Pre-test Mean (SD) | Post-test Mean (SD) | Change Mean (95% CI)[1] |
|---|---|---|---|---|
| Graphical Representations | Pre-test | 71.6 (19.5) | 71.5 (21.5) | -0.1 (-2.3, 2.1) |
| | GPA | 56.2 (26.5) | 65.2 (23.5) | 8.9 (5.8, 12.0)*** |
| Data collection and design | Pre-test | 74.7 (20.9) | 74.6 (23.0) | -0.2 (-3.3, 2.9) |
| | GPA | 66.9 (22.6) | 74.5 (22.8) | 7.5 (3.6, 11.4)*** |
| Descriptive statistics | Pre-test | 57.0 (37.6) | 56.0 (37.8) | -0.1 (-5.3, 3.6) |
| | GPA | 45.2 (38.7) | 50.9 (39.7) | 5.7 (0.0, 11.1)* |
| Tests of significance | Pre-test | 63.6 (14.2) | 72.6 (16.5) | 9.1 (7.0, 11.1)*** |
| | GPA | 55.7 (16.6) | 70.5 (17.2) | 14.8 (12.1, 17.4)*** |
| Confidence Intervals | Pre-test | 52.7 (18.8) | 59.3 (21.0) | 6.9 (4.1, 9.7)*** |
| | GPA | 44.3 (20.8) | 58.1 (22.1) | 14.0 (10.4, 17.6)*** |
| Sampling Variability | Pre-test | 71.0 (31.1) | 69.0 (34.3) | -2.4 (-6.2, 1.5) |
| | GPA | 55.6 (37.5) | 60.7 (36.9) | 5.5 (0.1, 10.8)* |
| Probability/Simulation | Pre-test | 50.8 (34.6) | 56.8 (35.3) | 5.7 (1.6, 9.9)** |
| | GPA | 42.2 (35.9) | 53.5 (35.8) | 11.4 (5.9, 16.8)*** |

1. From a paired *t*-test comparing the pre-test and post-test scores